\documentclass[twocolumn]{jpsj3}
\usepackage{txfonts}
\usepackage{braket}
\newcommand{\ds}{\displaystyle}

\title{On the Hidden Order in URu$_{2}$Si$_{2}$ --- Antiferro Hexadecapole Order and its Consequences}

\def\runauthor{H. \textsc{Kusunose} and H. \textsc{Harima}}
\author{Hiroaki \textsc{Kusunose}\thanks{hk@ehime-u.ac.jp} and
Hisatomo \textsc{Harima}$^{1}$
}

\inst{Department of Physics, Ehime University, Matsuyama, Ehime 790-8577, Japan\\
$^{1}$Department of Physics, Graduate School of Science, Kobe University, Nada, Kobe 657-8501, Japan
}

\abst{
An antiferro ordering of an electric hexadecapole moment is discussed as a promising candidate for the long standing mystery of the hidden order phase in URu$_{2}$Si$_{2}$.
Based on localized $f$-electron picture, we discuss the rationale of the selected multipole and the consequences of the antiferro hexadecapole order of $xy(x^{2}-y^{2})$ symmetry.
The mean-field solutions and the collective excitations from them explain reasonably significant experimental observations: the strong anisotropy in the magnetic susceptibility, characteristic behavior of pressure versus magnetic field or temperature phase diagrams, disappearance of inelastic neutron-scattering intensity out of the hidden order phase, and insensitiveness of the NQR frequency at Ru-sites upon ordering.
A consistency with the strong anisotropy in the magnetic responses excludes all the multipoles in two-dimensional representations, such as $(O_{yz},O_{zx})$.
The expected azimuthal angle dependences of the resonant X-ray scattering amplitude are given.
The $(x^{2}-y^{2})$-type antiferro quadrupole should be induced by an in-plane magnetic field along $[110]$, which is reflected in the thermal expansion and the elastic constant of the transverse $(c_{11}-c_{12})/2$ mode.
The $(x^{2}-y^{2})$-type [$(xy)$-type] antiferro quadrupole is also induced by applying the uniaxial stress along $[110]$ direction [$[100]$ direction].
A detection of these induced antiferro quadrupoles under the in-plane magnetic field or the uniaxial stress using the resonant X-ray scattering provides a direct redundant test for the proposed order parameter.
}

\kword{multipole order, neutron scattering, resonant X-ray scattering, Holstein-Primakoff method}

\begin{document}
\maketitle

\section{Introduction}

The heavy-fermion superconductor URu$_{2}$Si$_{2}$\cite{Palstra85,Maple86,Baumann85} undergoes a phase transition into the mysterious phase referred to as ``Hidden Order'' (HO) at $T_{\rm o}=17.5$ K at ambient pressure.
The term, HO, is given by the fact that the phase transition is signaled clearly by a large anomaly in the specific heat, while its order parameter (OP) has not been identified by a number of experimental investigations over the last two decades.
In early stage of research, the observed tiny antiferro (AF) magnetic moment\cite{Broholm87,Broholm91} was considered as a key factor for the HO phase.
However, it turned out that it is the parasitic tiny moment, whose volume fraction is enhanced by applied pressure.\cite{Matsuda01,Niklowitz10}
The elastic neutron-scattering experiments under pressure revealed a sharp phase transition at $p_{\rm c}\sim1.5$ GPa, above which a type-I antiferromagnetic  (AFM) phase emerges with $10$-times increase of the magnetic moments ($\mu_{o}\sim0.4\mu_{\rm B}$) along $c$ axis (the ordering vector is $\mib{Q}=(100)$).\cite{Amitsuka99}
In contrast to the drastic increase of the magnetic moment across the $p$-induced phase transition, the $p$-dependence of the transition temperatures from the paramagnetic phase to the low-$p$ HO and the high-$p$ AFM phases is weak and continuous\cite{Hassinger08,Aoki09,Motoyama08} (the transition between two ordered phases is of the first order).
Note that the bulk superconductivity below $T_{\rm c}\sim 1.5$ K does not coexist with the high-$p$ AFM,\cite{Hassinger08} although similarity of the Fermi surface in both the HO and the AFM phases is pointed out by Shubnikov-de Haas measurements.\cite{Hassinger10}
The recent laser angle-resolved photoemission spectroscopy gives an evidence of the broken translational invariance in the HO phase.\cite{Yoshida10}

Meanwhile, it has developed that multipole moments of the localized $f$ electrons are regarded as a key concept to understand HO phases observed in various $f$-electron systems.\cite{Kuramoto09}
This concept should be useful when a long-range order is triggered predominantly by the localized component of $f$ electrons, even though URu$_{2}$Si$_{2}$ is one of the most spectacular examples showing the itinerant-localized duality inherent in the heavy fermions.\cite{Kuramoto90,Miyake90,Miyake91,Kuramoto92}
So far, the various rank of multipole orders have been proposed for the HO phase: the magnetic dipole,\cite{Nieuwenhuys87} the electric quadrupole,\cite{Ohkawa99,Santini94,Santini95,Santini98} the magnetic octupole,\cite{Kiss05} the electric hexadecapole\cite{Haule09,Haule10} and the magnetic dotriacontapole.\cite{Cricchio09}
In principle, the resonant X-ray scattering (RXS) can detect multipole orders up to the 4th rank,\cite{Lovesey05,Kusunose05} provided that a superlattice reflection from the E2 (electric quadrupole) transition has a measurable intensity.
The recent RXS study in a major region of $\mib{k}=(h0l)$ plane has observed no traces of significant intensity except from the AFM, indicating the absence of quadrupole orders within a searched $\mib{k}$ range including a conceivable nesting vector $\mib{Q}^{*}=(1.4,0,0)$.\cite{Amitsuka10}

Recently, Haule and Kotliar have proposed the hexadecapole scenario based on the electronic structure computation combined with the dynamical mean-field approximation,\cite{Haule09,Haule10} where the Kondo renormalization process is arrested by the competition with the crystalline electric field (CEF) splitting, yielding a self-induced OP across the CEF splitting.
Their scenario provides a natural explanation for similarity of two ordered phases, although they are characterized by distinct OPs.
It is also pointed out that the low-energy intensity seen by the inelastic neutron scattering only in the HO phase can be understood by the pseudo-Goldstone mode from almost energetically degenerate ordered phases.\cite{Haule10}

Moreover, a space group analysis has been carried out to narrow down possible candidates for the OP in the HO phase.\cite{Harima10}
The second order phase transition from $I4/mmm$ (No.~139) to $P42/mnm$ (No.~136) does not need any kind of lattice distortion, allowing the NQR frequency at Ru-site unchanged.
On the other hand, the AFH order corresponds to No.~128 ($P4/mnc$), which loses the four-fold axis at Ru-site, and it should affect the NQR frequency in contrast to the No.~136 order.
However, the asymmetry parameter of the field gradient at Ru-site might be quite small due to the higher-rank AFH order as will be shown in this paper.
Thus, no anomaly observed in the Ru-NQR frequency within experimental accuracy\cite{Saitoh05} gives no decisive choice among two plausible space groups.

Motivated by these circumstances, we first discuss the rationale of the selection of multipoles based on the localized $f$-electron picture and the group theoretical point of view.
The observed strong anisotropy in magnetic responses requires the singlet-singlet-singlet low-lying CEF model, among which the dipole, the hexadecapole and the dotriacontapole remain active.
We discuss that the hexadecapole is the most natural candidate for the HO, and then we elucidate the consequences of the AF hexadecapole (AFH) order of $xy(x^{2}-y^{2})$ symmetry in detail.
This paper is organized as follows.
In \S2, we classify the CEF basis and the multipoles under tetragonal $D_{4h}$ symmetry.
We narrow down relevant active multipoles, mainly according to consistency with the strong anisotropy in the magnetic susceptibility.
In \S3, we investigate the mean-field solutions and the collective excitations from them, which explain reasonably significant experimental observations.
The emphasis is placed on the induced quadrupole moment under the in-plane magnetic fields, which could provide a useful test for the proposed OP.
The final section summarizes the paper.

\section{Selection of Hidden Order Parameter Candidate}

\begin{table*}
\caption{The CEF states of $J=4$ multiplet under $D_{4h}$ symmetry. The CEF states are classified by the character of $C_{4}$ symmetry operation for later convenience, which is indicated in the parenthesis of the first column.}
\begin{center}
\begin{tabular}{clll}
\hline
Group & Symmetry & Wave function & Energy \\ \hline
$\ds s_{1}$ ($+1$) & $\ds \Ket{\Gamma_{1}^{(1)}}$ & $\ds \frac{1}{\sqrt{2}}\sin\theta\biggl(\Ket{+4}+\Ket{-4}\biggr)+\cos\theta\Ket{0}$ & $\ds \Delta_{1}+\frac{\Delta}{\cos2\theta}$ \\
        & $\ds \Ket{\Gamma_{1}^{(2)}}$ & $\ds \frac{1}{\sqrt{2}}\cos\theta\biggl(\Ket{+4}+\Ket{-4}\biggr)-\sin\theta\Ket{0}$ & $\ds \Delta_{1}-\frac{\Delta}{\cos2\theta}$ \\
        & $\ds \Ket{\Gamma_{2}}$ & $\ds \frac{1}{\sqrt{2}}\biggl(\Ket{+4}-\Ket{-4}\biggr)$ & $\Delta_{1}-\Delta$ \\ \hline
$\ds s_{2}$ ($-1$) & $\ds \Ket{\Gamma_{3}}$ & $\ds \frac{1}{\sqrt{2}}\biggl(\Ket{+2}+\Ket{-2}\biggr)$ & $\ds \Delta_{34}-\frac{\sqrt{7}}{14}\left[\sqrt{5}\Delta\tan2\theta-\Delta'\tan2\phi\right]$ \\
        & $\ds \Ket{\Gamma_{4}}$ & $\ds \frac{1}{\sqrt{2}}\biggl(\Ket{+2}-\Ket{-2}\biggr)$ & $\ds \Delta_{34}+\frac{\sqrt{7}}{14}\left[\sqrt{5}\Delta\tan2\theta-\Delta'\tan2\phi\right]$ \\ \hline
$\ds d$ ($0$)     & $\ds \Ket{\Gamma_{5}^{(1)}\pm}$ & $\ds \sin\phi\Ket{\pm3}+\cos\phi\Ket{\mp 1}$ & $\ds \Delta_{5}+\frac{\Delta'}{8\cos2\phi}$ \\
        & $\ds \Ket{\Gamma_{5}^{(2)}\pm}$ & $\ds \cos\phi\Ket{\pm3}-\sin\phi\Ket{\mp 1}$ & $\ds \Delta_{5}-\frac{\Delta'}{8\cos2\phi}$ \\ \hline
\end{tabular}
\end{center}
\label{cefbasis}
\end{table*}

In order to narrow down possible candidates of the hidden order parameter, let us classify the multipoles within $J=4$ multiplet in $f^{2}$ configuration.
Under the tetragonal ($D_{4h}$) symmetry, which is characterized by the CEF hamiltonian in terms of the Stevens operators,\cite{Kuramoto09,Kusunose08,Stevens52,Hutchings64}
\begin{equation}
H_{\rm CEF}=B_{20}O_{20}+B_{40}O_{40}+B_{44}O_{44}+B_{60}O_{60}+B_{64}O_{64},
\label{hcef}
\end{equation}
the CEF states are summarized in Table~\ref{cefbasis}.
Here, instead of the five CEF parameters $B_{mn}$, we use the center-of-mass of energies, $\Delta_{1}$, $\Delta_{34}$, $\Delta_{5}$ and the mixing parameters, $\theta$ ($0\le\theta\le\pi/2$) and $\phi$ ($0\le\phi\le\pi/2$) characterizing the CEF wave functions.
The relation between two sets of parameters is given by
\begin{align}
&B_{20}=-\frac{1}{42}(5\Delta_{1}+4\Delta_{34}+6\Delta_{5}),
\cr
&B_{40}=\frac{1}{18480}(29\Delta_{1}+12\Delta_{34}+4\Delta_{5}),
\cr
&B_{60}=\frac{1}{332640}(\Delta_{1}+8\Delta_{34}-12\Delta_{5}),
\cr
&B_{44}=\frac{1}{264\sqrt{35}}\Delta\tan2\theta+\frac{1}{528\sqrt{7}}\Delta'\tan2\phi,
\cr
&B_{64}=\frac{1}{792\sqrt{35}}\Delta\tan2\theta-\frac{1}{15840\sqrt{7}}\Delta'\tan2\phi,
\end{align}
where $\Delta=3\Delta_{1}+2\Delta_{34}+4\Delta_{5}$ and $\Delta'=23\Delta_{1}+16\Delta_{34}+12\Delta_{5}$ for notational simplicity.
For later convenience, the CEF states are classified into three groups, $s_{1}$ $(+1)$, $s_{2}$ ($-1$) and $d$ ($0$) by the character of $C_{4}$ operation (as indicated in the parenthesis).
Note that the groups $s_{1}$ and $d$ are independent with each other.

\begin{table*}
\caption{The definition of multipoles up to 5th rank under $D_{4h}$ symmetry, which are classified by the character of $C_{4}$ symmetry operation.}
\begin{center}
\begin{tabular}{ccll}
\hline
Group & Symmetry & Notation & Angle dependence \\ \hline
$\ds S_{1}$ ($+1$) & $\ds \Gamma_{1g}$ & $\ds O_{20}$ & $\ds \frac{1}{2}(3z^{2}-r^{2})$ \\
  &               & $\ds O_{40}$ & $\ds \frac{1}{8}(35z^{4}-30z^{2}r^{2}+3r^{4})$ \\
  &               & $\ds O_{44}$ & $\ds \frac{\sqrt{35}}{8}(x^{4}-6x^{2}y^{2}+y^{4})$ \\ \cline{2-4}
  & $\ds \Gamma_{1u}$ & $\ds D_{s4}$ & $\ds \frac{3\sqrt{35}}{2}xyz(x^{2}-y^{2})$ \\ \cline{2-4}
  & $\ds \Gamma_{2g}$ & $\ds H_{z}^{\alpha}$ & $\ds \frac{\sqrt{35}}{2}xy(x^{2}-y^{2})$ \\ \cline{2-4}
  & $\ds \Gamma_{2u}$ & $\ds J_{z}$ & $\ds z$ \\
  &               & $\ds T_{z}^{\alpha}$ & $\ds \frac{1}{2}z(5z^{2}-3r^{2})$ \\
  &               & $\ds D_{z}^{\alpha}$ & $\ds \frac{1}{8}z[63z^{4}-5(14z^{2}-3r^{2})r^{2}]$ \\
  &               & $\ds D_{z}^{\beta}$ & $\ds \frac{3\sqrt{35}}{8}z(x^{4}-6x^{2}y^{2}+y^{4})$ \\ \hline
$\ds S_{2}$ ($-1$) & $\ds \Gamma_{3g}$ & $\ds O_{22}$ & $\ds \frac{\sqrt{3}}{2}(x^{2}-y^{2})$ \\
  &               & $\ds O_{42}$ & $\ds \frac{\sqrt{5}}{4}(x^{2}-y^{2})(7z^{2}-r^{2})$ \\ \cline{2-4}
  & $\ds \Gamma_{3u}$ & $\ds T_{xyz}$ & $\ds \sqrt{15}xyz$ \\
  &               & $\ds D_{s2}$ & $\ds \frac{\sqrt{105}}{2}xyz(3z^{2}-r^{2})$ \\ \cline{2-4}
  & $\ds \Gamma_{4g}$ & $\ds O_{xy}$ & $\ds \sqrt{3}xy$ \\
  &               & $\ds H_{z}^{\beta}$ & $\ds \frac{\sqrt{5}}{2}xy(7z^{2}-r^{2})$ \\ \cline{2-4}
  & $\ds \Gamma_{4u}$ & $\ds T_{z}^{\beta}$ & $\ds \frac{\sqrt{15}}{2}z(x^{2}-y^{2})$ \\
  &               & $\ds D_{z}^{\gamma}$ & $\ds \frac{\sqrt{105}}{4}z(x^{2}-y^{2})(3z^{2}-r^{2})$ \\ \hline
$\ds D$ ($0$) & $\ds \Gamma_{5g}$ & $\ds O_{yz}$, $\ds O_{zx}$ & $\ds \sqrt{3}yz$, $\ds \sqrt{3}zx$ \\
  &               & $\ds H_{x}^{\alpha}$, $\ds H_{y}^{\alpha}$ & $\ds \frac{\sqrt{35}}{2}yz(y^{2}-z^{2})$, $\ds \frac{\sqrt{35}}{2}zx(z^{2}-x^{2})$ \\
  &               & $\ds H_{x}^{\beta}$, $\ds H_{y}^{\beta}$ & $\ds \frac{\sqrt{5}}{2}yz(7x^{2}-r^{2})$, $\ds \frac{\sqrt{5}}{2}zx(7y^{2}-r^{2})$ \\ \cline{2-4}
  & $\Gamma_{5u}$ & $\ds J_{x}$, $\ds J_{y}$ & $\ds x$, $\ds y$ \\
  &               & $\ds T_{x}^{\alpha}$, $\ds T_{y}^{\alpha}$ & $\ds \frac{1}{2}x(5x^{2}-3r^{2})$, $\ds \frac{1}{2}y(5y^{2}-3r^{2})$ \\
  &               & $\ds T_{x}^{\beta}$, $\ds T_{y}^{\beta}$ & $\ds \frac{\sqrt{15}}{2}x(y^{2}-z^{2})$, $\ds \frac{\sqrt{15}}{2}y(z^{2}-x^{2})$ \\
  &               & $\ds D_{x}^{\alpha}$, $\ds D_{y}^{\alpha}$ & $\ds \frac{1}{8}x[63x^{4}-5(14x^{2}-3r^{2})r^{2}]$, $\ds \frac{1}{8}y[63y^{4}-5(14y^{2}-3r^{2})r^{2}]$ \\
  &               & $\ds D_{x}^{\beta}$, $\ds D_{y}^{\beta}$ & $\ds \frac{3\sqrt{35}}{8}x(y^{4}-6y^{2}z^{2}+z^{4})$, $\ds \frac{3\sqrt{35}}{8}y(z^{4}-6z^{2}x^{2}+x^{4})$ \\
  &               & $\ds D_{x}^{\gamma}$, $\ds D_{y}^{\gamma}$ & $\ds \frac{\sqrt{105}}{4}x(y^{2}-z^{2})(3x^{2}-r^{2})$, $\ds \frac{\sqrt{105}}{4}y(z^{2}-x^{2})(3y^{2}-r^{2})$ \\ \hline
\end{tabular}
\end{center}
\label{multipoles}
\end{table*}

By means of the irreducible representation of the tetragonal point group, the multipoles up to 5th rank are defined as in Table~\ref{multipoles}.
The corresponding quantum operators are obtained by replacing the polynomial of $(x,y,z)$ with the symmetric product of $(J_{x},J_{y},J_{z})$.\cite{Kusunose08,Kuramoto09}
Here, we again classify the multipoles into three groups, $S_{1}$, $S_{2}$ and $D$ by the character of $C_{4}$ operation.
This classification is convenient to discuss the active multipoles with finite matrix elements among the classified CEF states.

\begin{table}
\caption{The active multipoles between the classified CEF states.}
\begin{center}
\begin{tabular}{c|ccc}
\hline
        & $s_{1}$ & $s_{2}$ & $d$ \\ \hline
$s_{1}$ & $S_{1}$     & $S_{2}$     & $D$ \\
$s_{2}$ & $S_{2}$     & $S_{1}$     & $D$ \\
$d$     & $D$     & $D$     & $S_{1}\oplus S_{2}$ \\ \hline
\end{tabular}
\end{center}
\label{multipolecef}
\end{table}

The active multipoles between the classified CEF states are summarized in Table~\ref{multipolecef}.
The multipoles having unit character of $C_{4}$ operation remain active in the ``diagonal'' parts, $s_{1}$ and $s_{1}$, or $s_{2}$ and $s_{2}$ in Table~\ref{multipolecef}.
Similarly, the ``off-diagonal'' part between $s_{1}$ and $s_{2}$ has the active multipoles with $-1$ character of $C_{4}$ operation.
The multipoles with $\pm 1$ character of $C_{4}$ operation remain in the diagonal part between $d$ and $d$.
The multipole operators belonging to the two-dimensional representation (zero character of $C_{4}$) appear in the ``off-diagonal'' part between ($s_{1}$, $s_{2}$) and $d$.

\begin{table}
\caption{The proposed models and the candidates of the order parameters.}
\begin{center}
\begin{tabular}{llll}
\hline
Model & CEF scheme & OPs & Group \\ \hline
Nieuwenhuys\cite{Nieuwenhuys87} & $\ds \Gamma_{1}^{(1)}$-$\ds \Gamma_{2}$-$\ds \Gamma_{1}^{(2)}$-$\ds \Gamma_{5}^{(1)}$ & $\ds J_{z}$ & $\ds S_{1}$ \\
Haule-Kotliar\cite{Haule09,Haule10} & $\ds \Gamma_{2}$-$\ds \Gamma_{1}^{(2)}$ & $\ds H_{z}^{\alpha}$, $\ds J_{z}$ & $\ds S_{1}$ \\
Ohkawa-Shimizu\cite{Ohkawa99} & $\ds \Gamma_{5}^{(1)}$ & $\ds J_{z}$, $\ds O_{22}$, $\ds O_{xy}$ & $\ds S_{1}\oplus S_{2}$ \\
Santini-Amoretti\cite{Santini94,Santini95,Santini98} & $\ds \Gamma_{3}\,(\Gamma_{4})$-$\ds \Gamma_{1}^{(1)}$-$\ds \Gamma_{2}$-$\ds \Gamma_{5}^{(1)}$ & $\ds O_{xy}\,(O_{22})$ & $\ds S_{2}$ \\
Kiss-Fazekas\cite{Kiss05} & $\ds \Gamma_{1}^{(1)}$-$\ds \Gamma_{4}$-$\ds \Gamma_{5}^{(1)}$-$\ds \Gamma_{2}$ & $\ds T_{z}^{\beta}$ & $\ds S_{2}$ \\
\hline
\end{tabular}
\end{center}
\label{propmodel}
\end{table}

With these preliminaries, let us examine the so far proposed OPs based on the multipole ordering models, which are summarized in Table~\ref{propmodel}.
According to the analyses of the magnetic susceptibility, $\chi(T)$, the most natural choice of the low-lying CEF scheme is $\Gamma_{1}^{(1)}$-$\Gamma_{2}$,\cite{Amoretti84} which is common in the related materials, UPt$_{2}$Si$_{2}$,\cite{Nieuwenhuys87} PrRu$_{2}$Si$_{2}$\cite{Mulders97,Michalski00} and their dilute systems.\cite{Morishita07}
Note that this choice for the low-lying CEF scheme is also reached by the electronic structure computation combined with the dynamical mean-field approximation ($\Gamma_{1}$ and $\Gamma_{2}$ are reversed due to the hybridization effect).\cite{Haule09}
A prominent feature of $\chi(T)$ in URu$_{2}$Si$_{2}$ as compared with the related materials is the strong anisotropy, namely, $\chi_{a}(T)$ shows the Van Vleck type almost $T$-independent behavior, while $\chi_{c}(T)$ exhibits the Curie-Weiss behavior at least below 300 K.
The strong anisotropy is observed in the magnetization process as well.\cite{Nieuwenhuys87}
This indicates that the excited $\Gamma_{5}$ doublets are located at least 300 K far from the low-lying CEF singlets, which makes $J_{x}$ and $J_{y}$ operators inactive at low temperatures.
The large energy separation between the singlets and the doublets results in all the multipoles in the two-dimensional representation (the group $D$) inactive, and hence they can be excluded for possible candidates of the hidden order parameter.

It should be noted that the observed $\chi_{c}(T)$ in URu$_{2}$Si$_{2}$ shows a peak structure at about $T^{*}\sim 40$ K.
This is incompatible with the simple analysis with the above $\Gamma_{1}^{(1)}$-$\Gamma_{2}$ CEF model, showing the Van Vleck saturation below the $\Gamma_{1}^{(1)}$-$\Gamma_{2}$ splitting.
Practically, the Fermi-liquid quasiparticle emerges below $T^{*}$ where the non-trivial inter-site effect should be taken into account beyond simple localized models.
In fact, the decreasing behavior of $\chi_{c}(T)$ below $T^{*}$ is diminished by Th, La and Y substitutions for U in URu$_{2}$Si$_{2}$.\cite{Amitsuka10a,Amitsuka92,Yokoyama02a}
Moreover, the comparable energy scale of $T^{*}$ and the low-lying CEF splitting is consistent with the fact that the dilute U alloys exhibit non-fermi-liquid behavior,\cite{Amitsuka93,Amitsuka94,Yokoyama02} which can be understood by the competition between the Kondo singlet and the CEF singlet ground states.\cite{Yotsuhashi02,Toth10}

The models proposed by Nieuwenhuys,\cite{Nieuwenhuys87} and Haule and Kotliar\cite{Haule09} are consistent with the susceptibility analyses.
On the other hand, the model by Ohkawa and Shimizu\cite{Ohkawa99} could explain solely the strong anisotropy in $\chi(T)$, provided that the rest of the CEF states are far from the ground doublet.
In this case, however, we would have to consider that URu$_{2}$Si$_{2}$ is an exceptional in the series of compounds.
Note also that no indications of the quadrupole orders predicted by this model have been observed especially by using the RXS measurement.\cite{Amitsuka10,Nagao05}
In the model by Santini and Amoretti,\cite{Santini94,Santini95,Santini98} it is assumed the large exchange coupling constant for their quadrupole order, $O_{xy}$ or $O_{22}$ and the possible $J_{z}$ order.
This assumption involves much higher CEF states, which results in many active multipoles including $J_{x}$ and $J_{y}$.
Consequently, the strong anisotropy in $\chi(T)$ disappears.
The model by Kiss and Fazekas\cite{Kiss05} has a similar difficulty where the low-lying singlets and the doublet in their model make $J_{x}$ and $J_{y}$ active.
In their octupole ordering model, the crossing of the singlet and the doublet states is indispensable for their main conclusions in the magnetization process and the $p$-induced transition to the AFM phase.

\begin{figure*}[tb]
\begin{center}
\includegraphics[width=15cm]{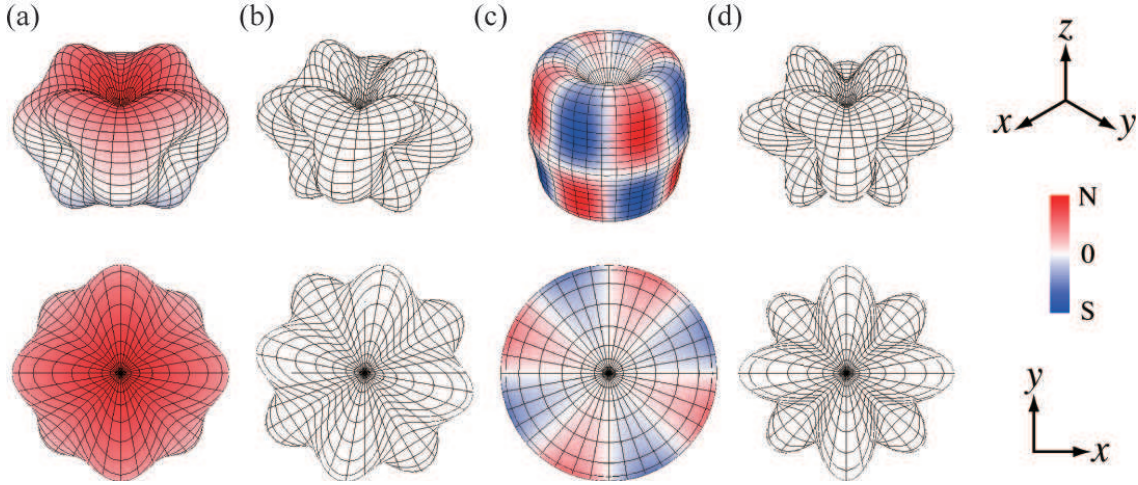}
\end{center}
\caption{(Color online) The local wave functions under (a) the AFM order, (b) the AFH order, (c) the AF dotriacontapole order, and (d) the paramagnetic state. The shape and the color-map represent the charge and the magnetic-charge densities, respectively.}
\label{wf}
\end{figure*}

Eventually, the models by Nieuwenhuys,\cite{Nieuwenhuys87} and Haule and Kotliar\cite{Haule09} could provide the most plausible candidate for the HO in view of the CEF scheme.
When we take the group $s_{1}$ as the low-lying CEF states, the multipoles in the group $S_{1}$ are natural candidates for the hidden OP occurring at low temperatures.
Apart from the possibilities of staggered ``scalar'' ($\Gamma_{1g}$) orders as was observed in PrRu$_{4}$P$_{12}$,\cite{Kuramoto09} the possibilities of $\Gamma_{2g}$-type hexadecapole ($H_{z}^{\alpha}$) and the $\Gamma_{1u}$-type dotriacontapole ($D_{s4}$) orders remain (the $\Gamma_{2u}$-type dipole $J_{z}$ is of course for the high-$p$ AFM phase).
The local wave functions under $J_{z}$, $H_{z}^{\alpha}$ and $D_{s4}$ orders are schematically shown in Fig.~\ref{wf}, where the shape and the color-map represent the charge and the magnetic-charge densities, respectively.\cite{Kusunose08}
The similarity of the local wave functions of Fig.~\ref{wf}(a) and (b) indicates slight change of whole electronic states across the $p$-induced phase boundary between the AFH and the AFM.
As was pointed out by Haule and Kotliar,\cite{Haule09,Haule10} the AFH and the AFM scenario explains other prominent experimental observations.\cite{Haule10}
In the next section, we discuss the consequences of these two orders in detail.

\section{AF Hexadecapole Order and its Consequences}

\subsection{The phase diagram based on $\Gamma_{1}^{(1)}$-$\Gamma_{2}$-$\Gamma_{1}^{(2)}$ CEF model}

Let us restrict ourselves to the low-lying CEF states in the group $s_{1}$.
Within the $\Gamma_{1}^{(1)}$-$\Gamma_{2}$-$\Gamma_{1}^{(2)}$ CEF model, the matrix elements of $J_{z}=-4\sigma$, $H_{z}^{\alpha}=105\xi$ and $D_{s4}=-630\eta$ are expressed as
\begin{align}
&\sigma=
\begin{pmatrix}
0 & \sin\theta & 0 \\
\sin\theta & 0 & \cos\theta \\
0 & \cos\theta & 0
\end{pmatrix},
\label{dmat}
\\
&\xi=
\begin{pmatrix}
0 & -i\cos\theta & 0 \\
i\cos\theta & 0 & -i\sin\theta \\
0 & i\sin\theta & 0
\end{pmatrix},\\
&\eta=
\begin{pmatrix}
0 & 0 & -i \\
0 & 0 & 0 \\
i & 0 & 0
\end{pmatrix}.
\label{emat}
\end{align}
To be specific, we set the CEF energies as $E_{1}^{(1)}=0$, $E_{2}=50$ K and $E_{1}^{(2)}=170$ K, which are roughly the same values as was used in the susceptibility analysis.\cite{Nieuwenhuys87}
The mixing parameter of the CEF wave functions is given by $\theta=0.998$.

First, we discuss the AFH and the AFM mean-field solutions based on the nearest-neighbor exchange model under external magnetic field $h_{z}=-4g_{J}\mu_{\rm B}H$ ($g_{J}=4/5$ in the $LS$-coupling scheme),
\begin{equation}
H=\frac{J}{z}\sum_{\Braket{i,j}}\sigma_{i}\sigma_{j}+\frac{D}{z}\sum_{\Braket{i,j}}\xi_{i}\xi_{j}-h_{z}\sum_{i}\sigma_{i}+\sum_{i}H^{\rm CEF}_{i},
\end{equation}
where $z=8$ is the coordination number and $H_{i}^{\rm CEF}$ is the diagonal matrix with $(E_{1}^{(1)},E_{2},E_{1}^{(2)})$.
The comment on the dotriacontapole $\eta$ will be given later.

\begin{figure}[tb]
\begin{center}
\includegraphics[width=8.5cm]{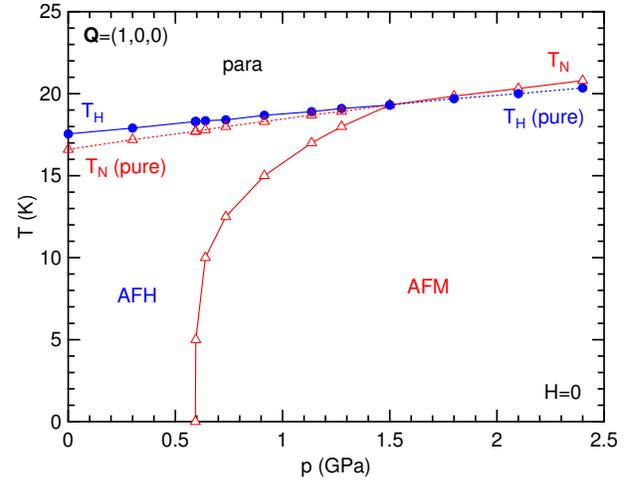}
\end{center}
\caption{(Color online) The mean-field $p$-$T$ phase diagram.
The dotted lines indicate the pure transition temperatures without other OPs. The boundary between the AFH and the AFM is of the first order.
}
\label{TpPhase}
\end{figure}

Assuming the linear $p$ dependences in $J$ and $D$ so as to reproduce the observed values of  $T_{\rm H}$ at ambient pressure, tri-critical point at $p_{\rm c}=1.5$ GPa and the zero-$T$ AFH-AFM boundary at $p=0.6$ GPa,\cite{Motoyama08} we obtain the mean-field $p$-$T$ phase diagram as shown in Fig.~\ref{TpPhase}.
Here, we have used $J(p)=J_{\rm c}+\alpha(p-p_{\rm c})$ and $D(p)=D_{\rm c}+\beta(p-p_{\rm c})$ with $J_{\rm c}=40.6$ K, $D_{\rm c}=91.4$ K, $\alpha=1.27$ K/GPa and $\beta=1.0$ K/GPa.

Since $\Gamma_{1}^{(2)}$ state is high enough as compared with the transition temperatures, the multipoles in the $\Gamma_{1}^{(1)}$-$\Gamma_{2}$ states are reduced to as $\sigma\to\sigma_{x}\sin\theta$, $\xi\to\sigma_{y}\cos\theta$ ($\mib{\sigma}$ is the Pauli matrix) and $\eta$ is vanishing.
$H^{\rm CEF}\to\sigma_{z}(E_{1}^{(1)}-E_{2})/2$ acts as the ``magnetic field'' along $z$ direction in the two low-lying CEF space.
This situation is discussed in the Ginzburg-Landau framework by Haule and Kotliar.\cite{Haule10}
Note that within the two low-lying CEF space the effective couplings at the tri-critical point have almost the same magnitude, i.e., $J_{c}\sin^{2}\theta\simeq D_{c}\cos^{2}\theta$ with the pseudo $U(1)$ (XY) symmetry.

In the AFH and the AFM phases, the local wave functions are represented by
\begin{align}
&
\Ket{\rm AFH}\sim\Ket{\Gamma_{1}^{(1)}}\pm i\alpha \Ket{\Gamma_{2}},
\\
&
\Ket{\rm AFM}\sim\Ket{\Gamma_{1}^{(1)}}\pm\alpha \Ket{\Gamma_{2}},
\end{align}
respectively, where $\alpha$ is a real constant depending on the temperature and the model parameters and the upper and the lower signs correspond to $A$ and $B$ sublattices.
Since $\Ket{\pm4}\propto \exp{(\pm i4\varphi)}$ ($\varphi$ is the azimuthal angle) and $\Ket{\Gamma_{2}}$ is pure imaginary, $\Ket{\rm AFH}$ is a real function that breaks the mirror symmetry with respect to the plane parallel to the $c$-axis in the electric potential, while $\Ket{\rm AFM}$ is a complex function representing a circular current with keeping the local symmetry in the electric potential.
In the $j$-$j$ coupling scheme within $j=5/2$ multiplet,\cite{Kusunose05a} these states are expressed by
\begin{align}
&
\Ket{\rm AFH}\sim
\frac{\cos\theta}{\sqrt{14}}\biggl(\Ket{d_{5}}+3\Ket{d_{3}}+2\Ket{d_{1}}\biggr)
\cr&\quad
+\frac{\sin\theta}{\sqrt{2}}\biggl(\Ket{+4}+\Ket{-4}\biggr)\pm i\frac{\alpha}{\sqrt{2}}\biggl(\Ket{+4}-\Ket{-4}\biggr),
\\
&
\Ket{\rm AFM}\sim
\frac{\cos\theta}{\sqrt{14}}\biggl(\Ket{d_{5}}+3\Ket{d_{3}}+2\Ket{d_{1}}\biggr)
\cr&\quad
+\frac{\sin\theta}{\sqrt{2}}\biggl(\Ket{+4}+\Ket{-4}\biggr)\pm\frac{\alpha}{\sqrt{2}}\biggl(\Ket{+4}-\Ket{-4}\biggr),
\end{align}
where $\Ket{+4}=\Ket{+5/2}\otimes\Ket{+3/2}$, $\Ket{-4}=\Ket{-3/2}\otimes\Ket{-5/2}$ and $\Ket{d_{m}}=\Ket{+m/2}\otimes\Ket{-m/2}$.
From these expressions, two ordered phases are essentially stabilized by the overlap among $\Ket{+4}+\Ket{-4}$ and $\Ket{+4}-\Ket{-4}$, and the relative phase in the linear combination discriminates two distinct orders.

\begin{figure}[tb]
\begin{center}
\includegraphics[width=8.5cm]{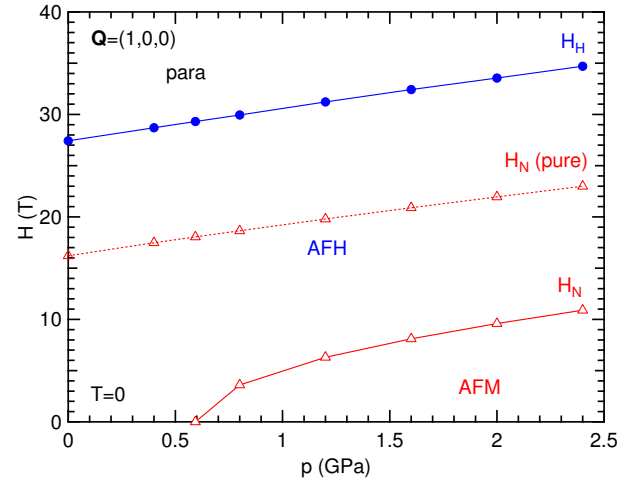}
\end{center}
\caption{(Color online) The mean-field $p$-$H$ ($H\parallel c$) phase diagram at $T=0$.}
\label{HpPhase}
\end{figure}

In Fig.~\ref{TpPhase}, the dotted lines indicate the pure transition temperatures without other OPs, which indicates that those orders are meta-stable states slightly higher than the real ordered states.
The 1st-order thermodynamic boundary between the AFH and the AFM phases is determined by the comparison of the free energies.
Note that the 1st-order phase boundary has a strong sample dependence\cite{Motoyama08,Aoki09,Amitsuka99} maybe due to uncontrollable inhomogeneity of samples under pressure.
This is consistent with the fact that even in the AFH phase the AFM exists, whose volume fraction increases by applied pressure.\cite{Matsuda01}
The suppression of the HO state due to the Rh impurities\cite{Yokoyama04} is discussed in the view of the competing orders of the AFM and the AFH.\cite{Pezzoli10}
With use of the above $J(p)$, $D(p)$ and the CEF parameters, the $p$-$H$ ($H\parallel c$) phase diagram at $T=0$ is obtained as shown in Fig.~\ref{HpPhase}.
The $p$ dependences of $H_{\rm H}$ and $H_{\rm N}$ are qualitatively agreement with the observed ones.\cite{Aoki09}

\subsection{The inelastic neutron scattering spectra}

Next, we discuss the inelastic neutron scattering spectra of the collective excitations from the above mean-field solutions.
A striking feature of the observed magnetic response is that both the CEF and the itinerant aspects are manifested.\cite{Broholm87,Broholm91}
Namely, the sharp CEF-like excitations propagate along the tetragonal basal plane at low energies, whereas fluctuations propagating along the $c$-axis constitute the high-energy broad magnetic excitations.
The dispersive excitations have a gap of 2 meV at the AF zone center and are strongly damped out above $T_{\rm o}$.
In the HO phase, the well-defined propagating magnon develops.
However, upon entering the AFM ($T_{\rm N}=1.5$ K) by decreasing $T$ at $p=0.67$ GPa, neither quasi-elastic nor inelastic responses at $\mib{Q}$ have been observed.\cite{Villaume08}
Thus, these excitations are characteristic of the HO phase.\cite{Villaume08,Yokoyama04,Aoki09}

Using the generalized Holstein-Primakoff framework,\cite{Kusunose01,Shiina03,Shiina04,Kusunose09} we calculate the dynamical structure function at $T=0$,
\begin{equation}
S(\mib{k},\epsilon)=-\frac{1}{\pi}{\rm Im}\,\chi_{z}(\mib{k},\epsilon),
\end{equation}
where $\chi_{z}(\mib{k},\epsilon)$ is the dipole-dipole green function defined as
\begin{equation}
\chi_{z}(\mib{k},\epsilon)=\Braket{\Braket{\sigma;\sigma}}_{\mib{k}}(\epsilon)=-i\int_{0}^{\infty}dt\,\Braket{\left[\sigma(\mib{k},t),\sigma(-\mib{k})\right]}e^{i\epsilon t}.
\end{equation}
The dynamical structure function of the hexadecapole scattering is also calculated in a similar manner with $\Braket{\Braket{\xi;\xi}}_{\mib{k}}(\epsilon)$, although it cannot be measured by present experimental techniques.

\begin{figure}[tb]
\begin{center}
\includegraphics[width=8.5cm]{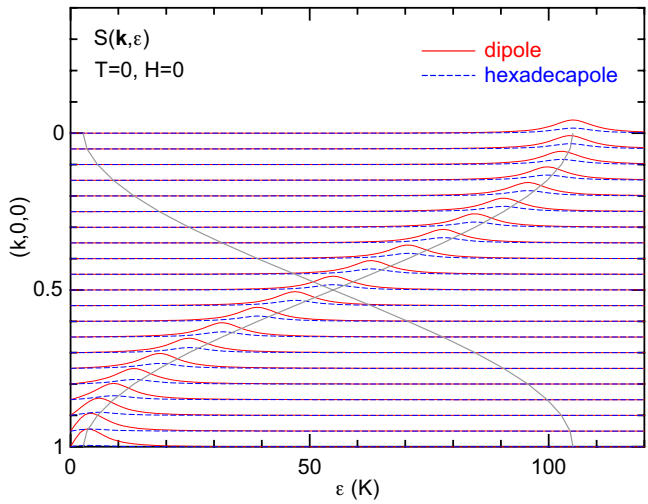}
\end{center}
\caption{(Color online) The inelastic neutron scattering spectra at ambient pressure (in the AFH phase) in the nearest-neighbor exchange model. The dynamical structure function for the hexadecapole scattering is also shown (the dashed lines). The gray lines indicate the dispersion relations for the collective excitations.}
\label{Skep0}
\end{figure}

\begin{figure}[tb]
\begin{center}
\includegraphics[width=8.5cm]{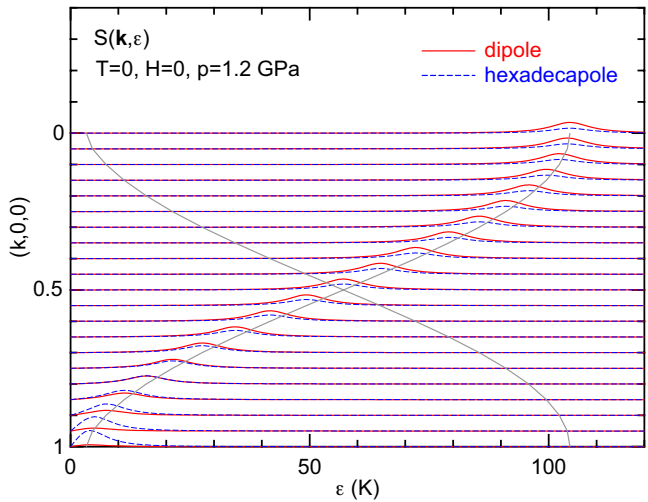}
\end{center}
\caption{(Color online) The inelastic neutron scattering spectra at $p=1.2$ GPa (in the AFM phase). The dipole intensity is diminished near the ordering vector $\mib{Q}=(1,0,0)$.}
\label{Skep12}
\end{figure}

Figure~\ref{Skep0} shows the excitation spectra at $T=0$ in the AFH phase at ambient pressure along $(100)$ direction.
The intensity due to the magnetic scattering slightly increases toward the ordering vector $\mib{Q}$, while the hexadecapole scattering intensity is rapidly decreasing toward $\mib{Q}$.
Similarly, the excitation spectra in the AFM phase at $p=1.2$ GPa is shown in Fig.~\ref{Skep12}.
In contrast to the AFH phase, the magnetic scattering intensity is almost diminished toward $\mib{Q}$, while the hexadecapole scattering remains finite.
In general, the collective mode is excited by transverse components (with off-diagonal matrix elements) with respect to the OP component.
In the AFH phase, the magnetic dipole $J_{z}$ plays a role of the transverse component, while it is the hexadecapole that plays this role in the AFM phase.
The latter is invisible by the neutron scattering.
Here, it is again important that the low-lying CEF states do not contain the active dipole operators $J_{x}$ and $J_{y}$.
This mechanism for the missing intensity in the AFM phase is also pointed out by Haule and Kotliar,\cite{Haule10} and a similar discussion was given in the context of the $\Gamma_{5}$ doublet CEF model.\cite{Ohkawa99}
Note that the missing intensity in the AFM phase should be recovered away from $\mib{Q}$ as shown in Fig.~\ref{Skep12}.
This provides a natural explanation of the mode observed in neutron scattering experiments under pressure.\cite{Broholm87,Broholm91,Villaume08,Yokoyama04,Aoki09}

\begin{figure}[tb]
\begin{center}
\includegraphics[width=8.5cm]{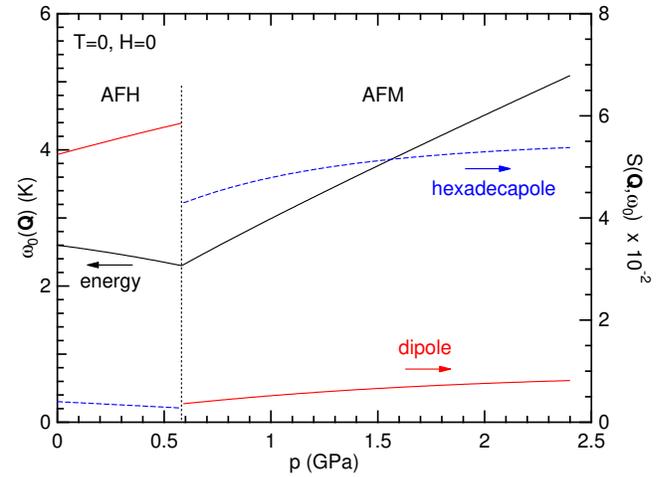}
\end{center}
\caption{(Color online) The $p$ dependence of the excitation energy and the intensities at $T=0$.}
\label{w0p}
\end{figure}

The $p$ dependence of the energy of the collective excitation at $\mib{Q}$ and its intensities is shown in Fig.~\ref{w0p}.
The increasing magnetic scattering intensity with increase of $p$ suddenly drops by an order of magnitude across the 1st-order transition to the AFM phase, at which the excitation energy $\omega_{0}(\mib{Q})$ takes a minimum value.
On the contrary, the invisible hexadecapole scattering is largely intensified across the phase boundary.

\begin{figure}[tb]
\begin{center}
\includegraphics[width=8.5cm]{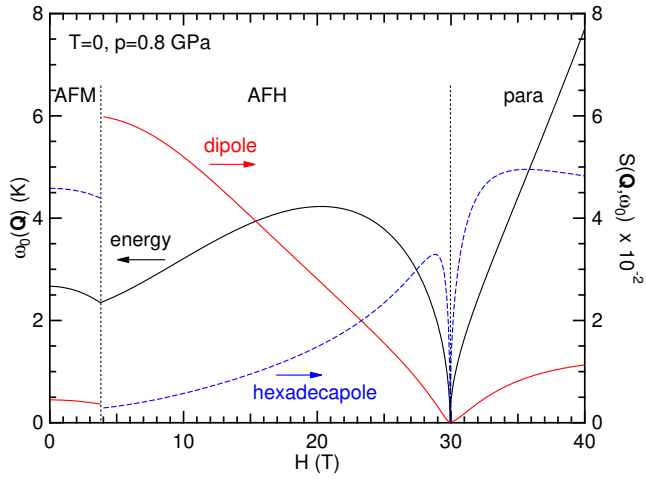}
\end{center}
\caption{(Color online) The $H\parallel c$ dependence of the excitation energy and the intensities at $p=0.8$ GPa and $T=0$.}
\label{w0H}
\end{figure}

Similarly, the $\mib{H}$ ($\parallel c$) dependence of the intensity is shown in Fig.~\ref{w0H}.
The excitation energy has a non-monotonous $H$ dependence in the AFH phase.
It is vanishing toward the second-order $H_{\rm H}$ since the magnitude of the OP is vanishing as well.
The excitation energy gap is roughly given by the energy difference of two ordered phases, and fluctuations around them constitute the pseudo Goldstone mode.\cite{Haule10}

At this point, we shall mention the possibility of the dotriacontapole order $\eta$.
Since the position of $\Gamma_{1}^{(2)}$ is unclear due to the lack of direct experimental information on the CEF scheme, one might consider that the AF $\eta$ order would take over the AFH order, provided that the $\Gamma_{1}^{(2)}$ state is low enough in energy.
However, since $\eta$ is connecting between two $\Gamma_{1}$'s while $\sigma$ is connecting between $\Gamma_{1}$ and $\Gamma_{2}$ (see eqs.~(\ref{dmat}) and (\ref{emat})), the occurrence of the AFM order has little influence on the $\eta$ operator and vice versa.
Hence the drastic change of the magnetic scattering intensity across the phase boundary does not occur in the case of the AF $\eta$ order.
This is confirmed by a similar calculation in the above (not shown).
By this reason, the AF dotriacontapole order is not a plausible candidate for the HO phase.

\subsection{The RXS amplitude and the field-induced moments}

\begin{table}
\caption{The angle dependence of the RXS {\it amplitude} in the nonrotated ($\sigma$-$\sigma'$) and the rotated ($\sigma$-$\pi'$) polarization channels both for the E1 and E2 transitions in the presence of the AFH order. The 3rd and the 4th rows indicated by $(O_{22})$ represent the angle dependence due to the induced AF $O_{22}$ order under the in-plane magnetic field.}
\begin{center}
\begin{tabular}{cccc}
\hline
$\mib{Q}$ & \multicolumn{2}{c}{transition} & angle dependence \\ \hline
 001 & E2 & $\sigma$-$\sigma'$   & $\ds \frac{1}{2\sqrt{2}}\cos^{2}\theta\sin4\psi$ \\
     &              & $\sigma$-$\pi'$      & $\ds -\frac{1}{\sqrt{2}}\sin\theta\cos^{2}\theta\cos4\psi$ \\ \hline
 100 & E2 & $\sigma$-$\sigma'$   & 0 \\
     &              & $\sigma$-$\pi'$      & $\ds \frac{\cos\theta}{16\sqrt{2}}\left[(3\cos2\theta-5)\sin\psi-2\cos^{2}\theta\sin3\psi)\right]$ \\ \hline
 001 & E1  & $\sigma$-$\sigma'$   & $\ds -\frac{1}{\sqrt{2}}\cos2\psi$ \\
 $(O_{22})$    &              & $\sigma$-$\pi'$      & $\ds -\frac{1}{\sqrt{2}}\sin\theta\sin2\psi$ \\ \hline
 100 & E1  & $\sigma$-$\sigma'$   & $\ds \frac{1}{2\sqrt{2}}(1-\cos2\psi)$ \\
 $(O_{22})$    &              & $\sigma$-$\pi'$      & $\ds -\frac{1}{2\sqrt{2}}\sin\theta\sin2\psi$ \\ \hline
\end{tabular}
\end{center}
\label{rxs}
\end{table}

Let us now consider direct consequences of the AFH order.
One of the powerful tools to observe the multipole order is the RXS measurement.\cite{Lovesey05}
Since the multipole order accompanies a specific shape in the charge and/or the magnetic-charge densities different from the original lattice symmetry, the azimuthal angle dependence of the RXS intensity provides a direct evidence for the OP symmetry.
According to a concise formalism for the RXS amplitude,\cite{Kusunose05} which assumes a spherical symmetry in the intermediate resonant processes of a core hole, the angle dependences of the scattering amplitude are summarized in Table~\ref{rxs}.
Here, $\theta$ is the scattering angle and $\psi$ is the azimuthal angle, whose origin is chosen when the $\sigma$-polarization vector $\mib{\epsilon}_{\sigma}$ is parallel to $a$ axis in the case of the rotation axis $\mib{Q}=(001)$.
The detailed experimental geometry of the scattering is shown in Fig.~1 of ref.~\citen{Kusunose05}.
Note that although $\mib{Q}=(001)$ and $(100)$ are essentially equivalent momenta in the Brillouin zone, the azimuthal dependences with respect to these two rotation axes are different.
The 3rd and 4th rows in Table~\ref{rxs} indicated by $(O_{22})$ will be explained below.

The presence of the AFH order causes an additional electric field which breaks the $D_{4h}$ symmetry at U-site.\cite{Kusunose08}
The additional electric potential at U-site is given by
\begin{equation}
\frac{\phi_{4}(\mib{r})}{-e}=\xi(\mib{Q})\frac{c}{r}\frac{\Braket{r^{4}}}{r^{4}}\frac{xy(x^{2}-y^{2})}{r^{4}},
\end{equation}
where $\braket{r^{4}}$ is the average of $r^{4}$ over the $5f$ radial wave function.
In the $LS$-coupling scheme, $c=-14\sqrt{35}/363$.
The resultant electric potentials generated at Si- and Ru-sites are proportional to $xy(x^{2}-y^{2})$ and $xy$, respectively.
This is consistent with the space group, No. 128 ($P4/mnc$) corresponding to the AFH order.\cite{Harima10}
The 4-fold axis at Ru-sites is lost by the $xy$-type electric potential, which should result in a finite asymmetry parameter $(\eta)$ in the electric field gradient tensor.
However, the NQR frequency at Ru-site has shown no changes at the onset of the HO within experimental accuracy.\cite{Saitoh05}
The lack of the frequency shift may be ascribed to smallness of the electric field due to the higher-rank hexadecapole order.
In fact, the ratio of $\phi_{4}(\mib{r})$ and the electric potential from the 2nd-rank $\Braket{O_{20}}$ in the CEF potential, $\phi_{2}(\mib{r})$, is roughly estimated as
\begin{equation}
\frac{\phi_{4}(\mib{r})}{\phi_{2}(\mib{r})}\sim 10^{-3}\times \frac{\Braket{r^{4}}}{r^{2}\Braket{r^{2}}},
\end{equation}
which is at least three order of magnitude small.
This smallness is also consistent with the fact that no measurable lattice distortions have been observed.

\begin{figure}[tb]
\begin{center}
\includegraphics[width=8.5cm]{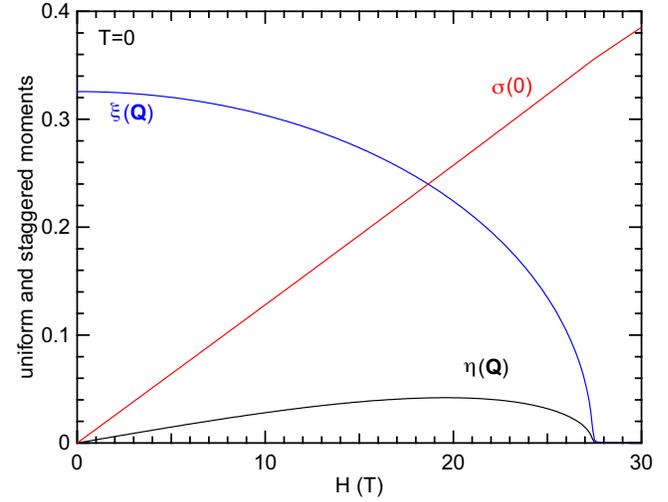}
\end{center}
\caption{(Color online) The $H\parallel c$ dependence of the ordered hexadecapole moment $\xi(\mib{Q})$ and the induced dipole and dotriacontapole moments.}
\label{usH}
\end{figure}

When the magnetic field along the $c$-axis is applied, the AF dotriacontapole is induced through the 3rd-order term,
\begin{equation}
{\cal F}_{3}\sim\sigma(0)\xi(\mib{Q})\eta(\mib{Q}),
\end{equation}
in the Ginzburg-Landau (GL) free energy.
The $H$ ($\parallel c$) dependence of $\sigma(0)$, $\xi(\mib{Q})$ and $\eta(\mib{Q})$ at $T=0$ are shown in Fig.~\ref{usH}.
The induced AF dotriacontapole has a maximum around the field at which $\sigma(0)\times \xi(\mib{Q})$ becomes the largest.
The induced $\eta(\mib{Q})$ causes an additional internal magnetic field along $c$-axis at Si-site, but not at Ru-site.
It is interesting to note that the single-crystal $^{29}$Si NMR line shows an additional broadening in the temperature range $[14,17.5]$ K,\cite{Takagi07} which may be relevant to the induced $\eta(\mib{Q})$.

\begin{figure}[tb]
\begin{center}
\includegraphics[width=8.5cm]{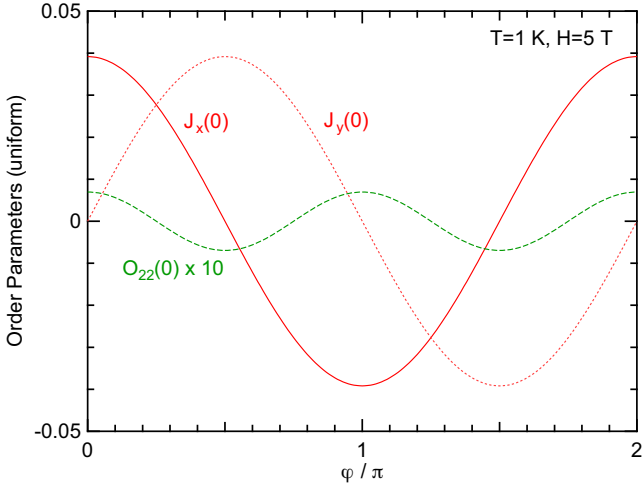}
\end{center}
\caption{(Color online) The in-plane magnetic-field angle dependence of the uniform ordered moments.}
\label{uHangle}
\end{figure}

\begin{figure}[tb]
\begin{center}
\includegraphics[width=8.5cm]{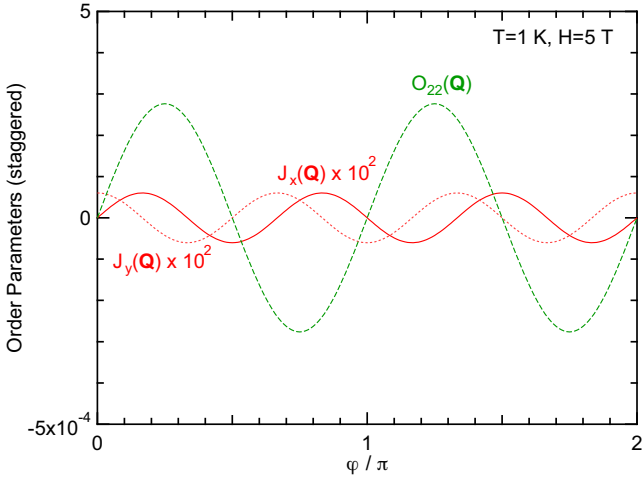}
\end{center}
\caption{(Color online) The in-plane magnetic-field angle dependence of the staggered ordered moments.}
\label{sHangle}
\end{figure}

We consider the effect of a magnetic field in the basal plane.
Since the GL free energy contains the 4th-order coupling,
\begin{equation}
{\cal F}_{4}\sim J_{x}(0)J_{y}(0)\xi(\mib{Q})O_{22}(\mib{Q}),
\end{equation}
the in-plane magnetic field should induce the AF $O_{22}$ ($x^{2}-y^{2}$)-type quadrupole order.
However, the induced moments are expected to be small since they are almost inactive at low temperatures.
In order to elucidate the induced orders semi-quantitatively, we adopt the mean-field theory with the full CEF states under eq.~(\ref{hcef}).
The in-plain field angle dependences [$\mib{H}=H(\cos\varphi,\sin\varphi,0)$] of the uniform and the staggered components of $J_{x}$, $J_{y}$ and $O_{22}$ are shown in Figs.~\ref{uHangle} and \ref{sHangle}.
Here, we have used $H=5$ T at $T=1$ K, and the CEF parameters, $B_{20}=-5.18$, $B_{40}=-0.744$, $B_{44}=0.546$, $B_{60}=B_{64}=0$, which reproduce the low-lying CEF states in the $\Gamma_{1}^{(1)}$-$\Gamma_{2}$-$\Gamma_{1}^{(2)}$ model.
All of the induced dipole moments, $J_{x}(\mib{Q})$ and $J_{y}(\mib{Q})$, are extremely small because of their inactiveness at low temperatures.
The induced AF $O_{22}$ quadrupole moment becomes the largest for the applied magnetic field along $[110]$.
Note that the uniform quadrupole $O_{22}(0)$ is also induced, although it is quite small.
It should be stressed that the induced AF quadrupole $O_{22}(\mib{Q})$ is also detectable by the RXS.
It is worth performing the RXS under the in-plane field along $[110]$ even in case that the RXS intensity of the AFH itself is too small to detect.
The angle dependence of the scattering amplitude is summarized in the rows indicated by $(O_{22})$ in Table~\ref{rxs}.
The in-plane magnetic-field measurements on the elastic response in the transverse $(c_{11}-c_{12})/2$ mode and the thermal expansion (it has already been performed without magnetic fields,\cite{Kuwahara97} showing the intrinsic softening in this mode) are highly desirable.

Similarly, a uniaxial stress can also induce AF quadrupoles.
Namely, through the 3rd-order coupling in the GL free energy,
\begin{equation}
{\cal F}_{3}\sim O_{xy}(0)\xi(\mib{Q})O_{22}(\mib{Q}),
\,\,\,
\text{or}
\,\,\,
{\cal F}_{3}\sim O_{xy}(\mib{Q})\xi(\mib{Q})O_{22}(0),
\end{equation}
the AF quadrupoles of the $(x^{2}-y^{2})$-type and the $(xy)$-type are induced by applying the uniaxial stresses along $[110]$ and $[100]$ directions, which correspond to the strain field of $\epsilon_{xy}$ and $\epsilon_{v}=\epsilon_{xx}-\epsilon_{yy}$, respectively.
Thus, a detection of these induced AF quadrupole under the uniaxial stress also provides a useful test for the AF hexadecapole order, although the magnitude of the induced quadrupole moment may be very small due to inactiveness of these multipoles in low temperatures.

The presence of the induced AF $O_{22}$ quadrupole breaks the 4-fold symmetry at U-site.
Quite recently, the emergence of two-fold oscillations in the magnetic torque under in-plane field rotation are sensitively detected in small pure crystals.\cite{Okazaki11,Thalmeier11}
It should be stressed that the induced $O_{22}$ quadrupole with the two-fold symmetry does not necessarily result in the two-fold oscillation in the magnetic torque, since the induced magnetic and quadrupole moments themselves depend on the magnetic field direction.
Nevertheless, a relevance between the two-fold oscillations and the induced $O_{22}$ quadrupole might be interesting to address in future investigations.

\section{Summary and Outlook}

We have discussed the nature of the AF hexadecapole order and its consequences on the basis of the localized $f$-electron exchange model.
The hexadecapole moment has been selected as a plausible candidate for the hidden order parameter in view from the consistency with the strong anisotropy in the magnetic susceptibility and the magnetization process.
This consistency requires the large energy separation between the low-lying CEF singlets and the excited doublets, which excludes all the multipoles in two-dimensional representations from candidates for the hidden order parameter.

The mean-field and the collective-excitation analyses provide reasonable agreements with the characteristic of the phase diagram and the missing neutron-scattering intensity out of the hidden order phase.
No significant change in the Ru-NQR frequency upon entering the hidden order phase can be understood by smallness of the electric potential associated with this higher-rank order parameter.
Nevertheless, this statement is unambiguously justified by more elaborate theoretical and/or experimental arguments on the electric potentials quantitatively.
The magnetic fields along the $c$-axis induces the AF dotriacontapole moment, which brings about additional internal magnetic fields at Si-sites of the same symmetry as the dipole ones.
Similarly, the AF $O_{22}$ quadrupole moment should be induced by the in-plane magnetic fields, which affects the thermal expansion and the elastic constant of the transverse $(c_{11}-c_{12})/2$ mode.
The uniaxial stress along $[110]$ and $[100]$ directions can also induce the AF $O_{22}$ and $O_{xy}$ quadrupoles, respectively.
It would be interesting to detect these induced AF quadrupoles by the RXS under the in-plane magnetic field or the uniaxial stress.

From general group-theoretical point of view, it is allowed a possibility that an order parameter in the high-$p$ AFM phase belongs to the subgroup of No.~128 space group corresponding to the low-$p$ AFH phase.
In this case, the high-$p$ phase is characterized by the coexistence of the AFM and the AFH, and the phase transition between the low-$p$ and the high-$p$ phases can be the second order.
Although the present work based on the localized exchange model does not have such a coexistent-phase solution, this possibility can provide a natural explanation for the similarity of the Fermi surfaces across the pressure-induced phase transition.

Since the superconductivity does not coexist with the AFM phase, it is presumable that the pairing is mediated by the pseudo Goldstone mode via the dipole coupling in the non-magnetic hexadecapole ordered state.\cite{Ishii95,Matsumoto04,Sato01}
The marked anisotropy in the upper critical field $H_{c2}$ in spite of rather 3-dimensional character of fermi surfaces\cite{Palstra85,Ohkuni99} can reconcile with the scenario of the magnetic exciton mediated superconductivity, which is sensitively damped against the magnetic field along $c$ axis but is insensitive for the in-plane fields (see Fig.~\ref{w0H} for example).
Thus, it would be interesting to elucidate an interplay between the magnetic excitations and the superconductivity in future investigations.

\section*{Acknowledgments}
This work was supported by a Grant-in-Aid for Scientific Research on Innovative Areas ``Heavy Electrons" (No.20102008) of The Ministry of Education, Culture, Sports, Science, and Technology (MEXT), Japan.
We would like to thank Hiroshi Amitsuka, Makoto Yokoyama, Tatsuya Yanagisawa, Hiroaki Ikeda, Takeshi Mito, Jacques Flouquet and Masashige Matsumoto for fruitful discussions.


\begin{thebibliography}{99}
\bibitem{Palstra85} T.T.M. Palstra, A.A. Menovsky, J. van der Berg, A.J. Dirkmaat, P.H. Kes, G.J. Nieuwenhuys and J.A. Mydosh: Phys. Rev. Lett. {\bf 55} (1985) 2727.
\bibitem{Maple86} M.B. Maple, J.W. Chen, Y. Dalichaouch, T. Kohara, C. Rossel and M.S. Torikachvili: Phys. Rev. Lett. {\bf 56} (1986) 185.
\bibitem{Baumann85} J. Baumann: Ph.D. thesis, University of Cologne (1985).
\bibitem{Broholm87} C. Broholm, J.K. Kjems, W.J.L. Buyers, P. Matthews, T.T. Palstra, A.A. Menovsky and J.A. Mydosh: Phy. Rev. Lett. {\bf 58} (1987) 1467.
\bibitem{Broholm91} C. Broholm, H. Lin, P.T. Matthews, T.E. Mason, W.J.L. Buyers, M.F. Collins, A.A. Menovsky, J.A. Mydosh and J.K. Kjems: Phys. Rev. B {\bf 43} (1991) 12809.
\bibitem{Matsuda01} K. Matsuda, Y. Kohori, T. Kohara, K. Kuwahara and H. Amitsuka: Phys. Rev. Lett. {\bf 87} (2001) 087203.
\bibitem{Niklowitz10} P.G. Niklowitz, C. Pfleiderer, T. Keller, M. Vojta, Y.-K. Huang and J.A. Mydosh: Phys. Rev. Lett. {\bf 104} (2010) 106406.
\bibitem{Amitsuka99} H. Amitsuka, M. Sato, N. Metoki, M. Yokoyama, K. Kuwahara, T. Sakakibara, H. Morimoto, S. Kawarazaki, Y. Miyako and J.A. Mydosh: Phys. Rev. Lett. {\bf 83} (1999) 5114.
\bibitem{Hassinger08} E. Hassinger, G. Knebel, K. Izawa, P. Lejay, B. Salce and J. Flouquet: Phys. Rev. B {\bf 77} (2008) 115117.
\bibitem{Aoki09} D. Aoki, F. Bourdarot, E. Hassinger, G. Knebel, A. Miyake, S. Raymond, V. Taufour and J. Flouquet: J. Phys. Soc. Jpn. {\bf 78} (2009) 053701.
\bibitem{Motoyama08} G. Motoyama, H. Yokoyama, A. Sumiyama and Y. Oda: J. Phys. Soc. Jpn. {\bf 77} (2008) 123710.
\bibitem{Hassinger10} E. Hassinger, G. Knebel, T.D. Matsuda, D. Aoki, V. Taufour and J. Flouquet: Phys. Rev. Lett. {\bf 105} (2010) 216409.
\bibitem{Yoshida10} R. Yoshida, Y. Nakamura, M. Fukui, Y. Haga, E. Yamamoto, Y. \=Onuki, M. Okawa, S. Shin, M. Hirai, Y. Muraoka and T. Yokoya: Phys. Rev. B {\bf 82} (2010) 205108.
\bibitem{Kuramoto09} See for example, Y. Kuramoto, H. Kusunose and A. Kiss: J. Phys. Soc. Jpn. {\bf 78} (2009) 072001.
\bibitem{Kuramoto90} Y. Kuramoto and K. Miyake: J. Phys. Soc. Jpn. {\bf 59} (1990) 2831.
\bibitem{Miyake90} K. Miyake and Y. Kuramoto: J. Magn. Magn. Mater. {\bf 90}\&{\bf 91} (1990) 438.
\bibitem{Miyake91} K. Miyake and Y. Kuramoto: Physica B {\bf 171} (1991) 20.
\bibitem{Kuramoto92} Y. Kuramoto and K. Miyake: Prog. Theor. Phys. Suppl. {\bf 108} (1992) 199.
\bibitem{Nieuwenhuys87} G.J. Nieuwenhuys: Phys. Rev. B {\bf 35} (1987) 5260.
\bibitem{Ohkawa99} F.J. Ohkawa and H. Shimizu: J. Phys. Condens. Matter {\bf 11} (1999) L519.
\bibitem{Santini94} P. Santini and G. Amoretti: Phys. Rev. Lett. {\bf 73} (1994) 1027.
\bibitem{Santini95} P. Santini and G. Amoretti: Phys. Rev. Lett. {\bf 74} (1995) 4098.
\bibitem{Santini98} P. Santini: Phys. Rev. B {\bf 57} (1998) 5191.
\bibitem{Kiss05} A. Kiss and P. Fazekas: Phys. Rev. B {\bf 71} (2005) 054415.
\bibitem{Haule09} K. Haule and G. Kotliar: Nature Phys. {\bf 5} (2009) 796.
\bibitem{Haule10} K. Haule and G. Kotliar: Europhys. Lett. {\bf 89} (2010) 57006.
\bibitem{Cricchio09} F. Cricchio, F. Bultmark, O. Gr\r{a}n\"as and L. Nordstr\"om: Phys. Rev. Lett. {\bf 103} (2009) 107202 (in which the single-particle classification for the multipole is used).
\bibitem{Lovesey05} See for example, S.W. Lovesey, E. Balcar, K.S. Knight and J. Fem\'andez Rodriguez: Phys. Rep. {\bf 411} (2005) 233.
\bibitem{Kusunose05} H. Kusunose and Y. Kuramoto: J. Phys. Soc. Jpn. {\bf 74} (2005) 3139.
\bibitem{Amitsuka10} H. Amitsuka, T. Inami, M. Yokoyama, S. Takayama, Y. Ikeda, I. Kawasaki, Y. Homma, H. Hidaka and T. Yanagisawa: to be published in J. Phys.: Conf. Series {\bf 200} (2010) 012007.
\bibitem{Nagao05} T. Nagao and J. Igarashi: J. Phys. Soc. Jpn. {\bf 74} (2005) 765.
\bibitem{Harima10} H. Harima, K. Miyake and J. Flouquet: J. Phys. Soc. Jpn. {\bf 79} (2010) 033705.
\bibitem{Saitoh05} S. Saitoh, S. Takagi, M. Yokoyama and H. Amitsuka: J. Phys. Soc. Jpn. {\bf 74} (2005) 2209.
\bibitem{Kusunose08} H. Kusunose: J. Phys. Soc. Jpn. {\bf 77} (2008) 064710.
\bibitem{Stevens52} K.W.H. Stevens: Proc. Phys. Soc., Sect. A{\bf 65} (1952) 209.
\bibitem{Hutchings64} M.T. Hutchings: Solid State Phys. {\bf 16} (1964) 227.
\bibitem{Amoretti84} G. Amoretti, A. Blaise and J. Mulak: J. Magn. Magn. Mater. {\bf 42} (1984) 65.
\bibitem{Mulders97} A.M. Mulders, A. Yaouanc, P.D. de R\'eotier, P.C.M. Gubbens, A.A. Moolenaar, B. F\r{a}k, E. Ressouche, K. Proke\v{s}, A.A. Menovsky and K.H.J. Buschow: Phys. Rev. B {\bf 56} (1997) 8752.
\bibitem{Michalski00} R. Michalski, Z. Ropka and R.J. Radwanski: J. Phys.: Condens. Matter {\bf 12} (2000) 7609.
\bibitem{Morishita07} A. Morishita, K. Matsuda, T. Wakabayashi, I. Kawasaki, K. Tenya and H. Amitsuka: J. Mag. Mag. Mater. {\bf 310} (2007) 283.
\bibitem{Amitsuka10a} H. Amitsuka and M. Yokoyama: private communication.
\bibitem{Amitsuka92} H. Amitsuka, K. Tateyama, C.C. Paulsen, T. Sakakibara and Y. Miyako: J. Mag. Mag. Mater. {\bf 104}-{\bf 107} (1992) 60.
\bibitem{Yokoyama02a} M. Yokoyama, K. Tenya and H. Amitsuka: Physica B {\bf 312}-{\bf 313} (2002) 498.
\bibitem{Amitsuka93} H. Amitsuka, T. Hidano, T. Honma, H. Mitamura and T. Sakakibara: Physica B {\bf 186-188} (1993) 337.
\bibitem{Amitsuka94} H. Amitsuka and T. Sakakibara: J. Phys. Soc. Jpn. {\bf 63} (1994) 736.
\bibitem{Yokoyama02} M. Yokoyama, H. Amitsuka, K. Kuwahara, K. Tenya and T. Sakakibara: J. Phys. Soc. Jpn. {\bf 71} (2002) 3037.
\bibitem{Yotsuhashi02} S. Yotsuhashi, K. Miyake and H. Kusunose: J. Phys. Soc. Jpn. {\bf 71} (2002) 389.
\bibitem{Kusunose05a} Appendix A in H. Kusunose and H. Ikeda: J. Phys. Soc. Jpn. {\bf 74} (2005) 405.
\bibitem{Toth10} A. T\'oth, P. Chandra, P. Coleman, G. Kotliar and H. Amitsuka: Phys. Rev. B {\bf 82} (2010) 235116.
\bibitem{Yokoyama04} M. Yokoyama, H. Amitsuka, S. Itoh, I. Kawasaki, K. Tenya and H. Yoshizawa: J. Phys. Soc. Jpn. {\bf 73} (2004) 545.
\bibitem{Pezzoli10} M.E. Pezzoli, M.J. Graf, K. Haule, G. Kotliar and A.V. Balatsky: arXiv:1012.3200.
\bibitem{Villaume08} A. Villaume, F. Bourdarot, E. Hassinger, S. Raymond, V. Taufour, D. Aoki and J. Flouquet: Phys. Rev. B {\bf 78} (2008) 012504.
\bibitem{Kusunose01} H. Kusunose and Y. Kuramoto: J. Phys. Soc. Jpn. {\bf 70} (2001) 3076.
\bibitem{Shiina03} R. Shiina, H. Shiba, P. Thalmeier, A. Takahashi and O. Sakai: J. Phys. Soc. Jpn. {\bf 72} (2003) 1216.
\bibitem{Shiina04} R. Shiina, M. Matsumoto and M. Koga: J. Phys. Soc. Jpn. {\bf 73} (2004) 3453.
\bibitem{Kusunose09} H. Kusunose, M. Matsumoto and M. Koga: J. Phys. Soc. Jpn. {\bf 78} (2009) 094713.
\bibitem{Takagi07} S. Takagi, S. Ishihara, S. Saitoh, H. Sasaki, H. Tanida, M. Yokoyama and H. Amitsuka: J. Phys. Soc. Jpn. {\bf 76} (2007) 033708.
\bibitem{Kuwahara97} K. Kuwahara, H. Amitsuka, T. Sakakibara, O. Suzuki, S. Nakamura, T. Goto, M. Mihalik, A.A. Menovsky, A. de Visser and J.J.M. Franse: J. Phys. Soc. Jpn. {\bf 66} (1997) 3251.
\bibitem{Okazaki11} R. Okazaki, T. Shibauchi, H.J. Shi, Y. Haga, T.D. Matsuda, E. Yamamoto, Y. Onuki, H. Ikeda and Y. Matsuda: Science {\bf 331} (2011) 439.
\bibitem{Thalmeier11} P. Thalmeier and T. Takimoto: arXiv:1102.3399.
\bibitem{Ishii95} H. Ishii, H. Ohgaki and A. Oguri: Phys. Rev. B {\bf 52} (1995) 12969.
\bibitem{Matsumoto04} M. Matsumoto and M. Koga: J. Phys. Soc. Jpn. {\bf 73} (2004) 1135.
\bibitem{Sato01} N.K. Sato, N. Aso, K. Miyake, R. Shiina, P. Thalmeier, G. Varelogiannis, C. Geibel, F. Steglich, P. Flude and T. Komatsubara: Nature {\bf 410} (2001) 340.
\bibitem{Ohkuni99} H. Ohkuni, Y. Inada, Y. Tokiwa, K. Sakurai, R. Settai, T. Honma, Y. Haga, E. Yamamoto, Y. \={O}nuki, H. Yamagami, S. Takahashi and T. Yanagisawa: Phil. Mag. B {\bf 79} (1999) 1045.
\end{thebibliography}
\end{document}